\documentclass[aps,preprint,showpacs,showkeys]{revtex4}
\usepackage{graphicx}
\begin{document}
\newcommand{\be}{\begin{equation}}
\newcommand{\ee}{\end{equation}}
\newcommand{\bea}{\begin{eqnarray}}
\newcommand{\eea}{\end{eqnarray}}
\title{A geometric approach to the thermodynamics of the van der Waals system}

\author{Hernando Quevedo$^{1,2,3}$ and Antonio Ram\'\i rez$^1$}

\affiliation{
$^1$Instituto de Ciencias Nucleares, 
Universidad Nacional Aut\'onoma de M\'exico, 
 AP 70543, M\'exico, DF 04510, Mexico\\ 
$^2$Dipartimento di Fisica and ICRA, Universit\'a di Roma ``La Sapienza", Piazzale Aldo Moro 5, I-00185 Roma, Italy\\
$^3$Department of Physics, Al-Farabi Kazakh National University, 050040 Almaty, Kazakhstan}

\begin{abstract}
We investigate the geometric properties of the equilibrium manifold of a thermodynamic system determined by the van der Waals equations of state.
We use the formalism of geometrothermodynamics to obtain results that are invariant under Legendre transformations, i. e., independent
of the choice of thermodynamic potential. It is shown that the equilibrium manifold is curved with curvature singularities situated 
at those points where first order phase transitions occur. Moreover, the geodesic equations in the equilibrium manifold are investigated numerically to illustrate the equivalence between geodesic incompleteness and curvature singularities as a criterion to define phase transitions in an invariant manner.
\end{abstract}
\relax

\maketitle

%%%%%%%%%%%%%%%%%%%%%%%%%%%%%%%%%%%%%%%%%%%%
%% MAINMATTER
%%%%%%%%%%%%%%%%%%%%%%%%%%%%%%%%%%%%%%%%%%%%

\section{Introduction}
\label{sec:int}
It is well known that in nature there exist four different interactions and all of them 
can be interpreted in terms of  concepts of differential geometry. 
Indeed, 
Einstein proposed the astonishing principle ``field strength = curvature" 
to understand the physics of the gravitational field (see, for instance, \cite{mtw,frankel}). 
In this case, 
curvature means the curvature of spacetime, a Riemannian manifold with a metric compatible with a connection that is unique 
as a consequence of the assumption that the torsion tensor vanishes. 
The second  element of general relativity are Einstein's field equations
$
R_{\mu\nu}- \frac{1}{2} g_{\mu\nu} R = 8\pi G\, T_{\mu\nu}
$
that established for the first time the amazing principle ``geometry = energy". 

The conceptual fundamentals of this principle were very controversial when first formulated; however, experimental 
evidence has shown its correctness and even modern generalizations of Einstein's theory follow the same principle. 
On the other hand, since the field strength can be considered as a measure of the gravitational interaction, we conclude 
that the entire idea of general relativity can be summarized in the principle ``interaction = curvature".

The discovery by Yang and Mills \cite{ym53} that the theory of electromagnetism can be interpreted as an Abelian gauge theory 
and that non-Abelian generalizations can be constructed in a similar way, represents a major achievement. Today, it is known that
electromagnetism  
can be described geometrically in terms of 
the elements of a principal fiber bundle. In this case, the base manifold is the Minkowski spacetime, the standard fiber is the gauge group $U(1)$, which 
represents the internal symmetry of electromagnetism,
and the connection across the fibers is a local cross-section that takes values in the algebra of $U(1)$.

This result opened the possibility of fixing the background metric, 
in accordance with the desired properties of the base manifold, and selecting different connections as local cross-sections 
of the principal fiber bundle. In particular, it was shown that connections with values in the Lie algebra of the gauge groups $SU(2)$ and $SU(3)$ 
can be used to represent the internal symmetries of the weak and strong interactions, respectively.
This interesting geometrical approach constituted the basis for the construction of modern 
gauge theories that are used as the starting point to formulate the physics of the electromagnetic, weak, and strong interactions. 
It follows that the principle ``curvature = interaction" holds for all known forces of nature, and is a fundamental element for the construction
of the corresponding gauge theories.  

Consider now the case of a thermodynamic system.
In very broad terms, one can say that in a thermodynamic system the constituents are subject to the action of all known forces.
However, it is not necessary to consider all the physical details of each interaction. In fact,  
  due to the large number of particles involved in the system, only a statistical approach is possible from which average values for
the physical quantities of interest, like energy, entropy, etc., are derived. 
The question arises whether it is also possible to find a geometric construction for which the 
principle ``curvature = thermodynamic interaction" holds. 
We will see in the present work that the formalism of geometrothermodynamics (GTD) \cite{quev07}
satisfies this requirement. First, we must mention that our interpretation of thermodynamic interaction is based upon the statistical approach 
to thermodynamics in which all  the properties of the system can be derived from the explicit form of the corresponding Hamiltonian \cite{greiner}. 
As usual, the interaction between the particles of the system is described by the potential part of the Hamiltonian. Consequently, if the potential vanishes, we say that the system does not possess thermodynamic interaction. 

In this work, we present the formalism of GTD using  Riemannian contact geometry for the definition of the thermodynamical phase 
manifold and the equilibrium manifold. It is explained why it is necessary to introduce the auxiliary phase manifold in order to take into account the Legendre transformations in the appropriate manner. We apply our formalism to study the geometric properties of the equilibrium manifold of the van der Waals gas,  and find that in general it is curved for arbitrary values of the parameters, except in the limiting case of an ideal gas for which the manifold turns out to be flat. 
Moreover, it is shown that the curvature singularities are located at those points where first order phase transitions occur. In addition, we explore the behavior of the geodesics, which represent quasi-static processes connecting different equilibrium states, and show that geodesic incompleteness can also be used as a criterion to detect points of phase transitions.

This paper is organized as follows. In Section \ref{sec:gtd} we introduce the main concepts of Riemannian contact geometry 
that are necessary to define the phase manifold and its equilibrium submanifolds. A particular metric is given that is invariant 
under total Legendre transformations, and is used to describe systems in which first order phase transitions can take place.
Section \ref{sec:equ} is dedicated to the investigation of 
the geometric properties of the equilibrium manifold for the van der Waals system, its curvature and singularities.  Section \ref{sec:geo} includes an analysis of the geodesic equations that are integrated numerically to find the points where geodesic incompleteness appears.
 Finally, Section \ref{sec:con} is devoted
to discussions of our results.
Throughout this paper we use units in which $k_{_B}=1$.

\section{Basic aspects of geometrothermodynamics}
\label{sec:gtd}

From a geometric point of view, the most important object of a given thermodynamic system is its thermodynamic space ${\cal E}$, each point of which represents an equilibrium state. To be more specific it is convenient to introduce coordinates, say $E^a$, on ${\cal E}$, where $a=1,2,\cdots, n$. The integer $n$ represents the number of thermodynamic degrees of freedom of the system. In general, it is possible to choose any set of $n$ coordinates for ${\cal E}$, but for the sake of convenience, the set $\{E^a\}$ of extensive thermodynamic variables is usually taken as the coordinates of the equilibrium space. Furthermore, a given thermodynamic system is uniquely characterized by the fundamental equation $\Phi=\Phi(E^a)$, where $\Phi$ is the thermodynamic potential that is used to describe the system \cite{callen}. 
The importance of the fundamental equation lies in the fact that from it one can derive all the equations of state that characterize the system in a unique manner. It follows that the fundamental equation can distinguish between different systems. 

Of course, for a fundamental equation to be physically relevant it must be in accordance with the laws of classical thermodynamics. For instance, the functional dependence of $\Phi$ must be such that the second law is satisfied. Moreover, the exterior derivative of the fundamental form (sum over repeated indices)
\begin{equation}
d\Phi = \frac{\partial \Phi}{\partial E^a} d E^a = I_a d E^a \ ,
\label{first}
\end{equation}
where $I_a$ represent the intensive variables dual to $E^a$, is equivalent to the first law of thermodynamics. 

With only the above information about the equilibrium space ${\cal E}$ one can just explore its topological properties and the analytic behavior of the fundamental equation $\Phi=\Phi(E^a)$. In some sense this is what has been done in analytic geometry in the last century. In fact, the fundamental equation determines a surface on ${\cal E}$ whose geometric properties are related to the thermodynamic behavior of the system. For instance, if the thermodynamic potential is chosen as the Gibbs potential it is known that the critical points of the surface can be used as a criterion to define phase transitions whose order is directly related to the order of the differential derivative of the Gibbs potential at which the critical points exist. The study of classical thermodynamics systems is based upon this definition of phase transitions and the investigation of sections of the equilibrium space. 

To extract more information from the equilibrium space it is necessary to equip it with an additional geometric structure. This can be done, for instance, by introducing a metric structure $g=g_{ab}dE^adE^b$ on ${\cal E}$. Since for a given thermodynamic system there exists only one fundamental equation $\Phi(E^a)$ that also determines the properties of ${\cal E}$, one would expect that the metric $g$ is determined in a unique manner by $\Phi(E^a)$. A particularly interesting choice is the Hessian metric
\begin{equation}
g^H_{ab} = \frac{\partial^2 \Phi}{\partial E^a E^b}  \ .
\label{hess}
\end{equation}
The specific choice with the entropy $S$ as the potential $\Phi$ can be considered as due originally to  
Rao \cite{rao45} who in 1945 proposed to use a metric structure in statistical physics, and constructed 
a particular metric whose components in local coordinates coincide with Fisher's
information matrix. Rao's original work has been followed up and extended by a number of
authors (see, e.g., \cite{amari85} for a review). Later on, in 1975, Weinhold \cite{wei75} proposed to use 
the internal energy $U$ as the thermodynamic potential. The corresponding metric can be shown to be flat for 
the fundamental equation of the ideal gas, and curved in other more general cases. The choice of the entropy as  
thermodynamic potential was proposed in 1979 by Ruppeiner \cite{rup79,rup95}, obtaining a metric which is 
conformally equivalent to Weinhold's metric,
with the inverse of the temperature as the conformal factor. 
The physical meaning of Ruppeiner's metric lays in the fluctuation
theory of equilibrium thermodynamics. It turns out that the second
moments of fluctuation are related to the components of
Ruppeiner's metric. More recently,  Liu,
L\"u, Luo and Shao \cite{chinosmet} proposed to use the potential $\Phi= 
\tilde U$, where $\tilde U$ is any of the
thermodynamic potentials that can be obtained from $U$ by means of a
Legendre transformation. In the special cases $\tilde U = U$ and
$\tilde U = S$, one obtains the Weinhold and Ruppeiner metrics,
respectively.

The pair $({\cal E}, g)$ is then a Riemannian manifold with specific geometric properties that are expected to be related to the thermodynamic properties of the system. As mentioned in the introduction, it is expected that the curvature of the equilibrium manifold be related to the notion of thermodynamic interaction. If this turns out to be true, it is then interesting to consider the thermodynamic interpretation of curvature singularities. Since in differential geometry the presence of curvature singularities is considered as indicating the break down of the geometric formalism, one can expect that such singularities are also related to the break down of the thermodynamic description. On the other hand, the limit of applicability of classical equilibrium thermodynamics coincide with the occurrence of phase transitions. We conclude that curvature singularities should correspond to phase transitions.

An additional important aspect is the fact that ordinary classical thermodynamics is invariant with respect to Legendre transformations, i. e., it is independent of the choice of thermodynamic potential \cite{callen}. Clearly, the geometric properties of the Hessian metrics defined above are not necessarily invariant with respect to Legendre transformations. In fact, since a Legendre transformation includes a change of potential $\Phi\rightarrow \tilde\Phi$ that is accompanied by an interchange of extensive and intensive variables, it could change the functional dependence of the thermodynamic potential $\Phi$, implying modifications of the explicit components of the metric (\ref{hess}). This change of functional dependence could in principle lead to a change of the geometric properties of the equilibrium manifold.  

One might then wonder whether the use of 
non Legendre invariant metric structures in ordinary thermodynamics 
would always lead to results that do not depend on the thermodynamic 
potential. Indeed, several examples are known in the literature 
in which a change of thermodynamic potential leads to a modification 
of the thermodynamic geometry \cite{scws07,mz07,med08,bc08,bc10,bc10a}. Some puzzling results and inconsistencies
arise also in connection with the use of different metrics in the 
equilibrium manifold \cite{abp06,sst08,bc11,wlwg10}, in the sense that for the same 
thermodynamic system the resulting geometry
can be either flat or curved, depending on the choice of thermodynamic potential chosen for generating the metric.

It then follows that the Legendre invariance of $g$ is necessary in order to guarantee in general that the geometric properties of the equilibrium manifold ${\cal E}$ do not depend on the choice of thermodynamic potential. GTD is a formalism that  incorporates Legendre invariance into the geometric structure of  ${\cal E}$ in a consistent manner. Since the Legendre transformations involve the thermodynamic potential as well as the extensive and the intensive thermodynamic variables, to introduce Legendre invariance it is necessary to consider an auxiliary space in which all the above variables are independent. This can be done by using the thermodynamic phase space ${\cal T}$ with coordinates $Z^A =\{\Phi,E^a,I^a\}$, where $I^a$ represent the intensive variables and $A=0,1,\cdots,2n$. Notice that a specific thermodynamic system cannot be considered in ${\cal T}$, but in the particular subspace where the fundamental equation $\Phi=\Phi(E^a)$ holds. 

A Legendre transformation can now be defined as a coordinate transformation in ${\cal T}$ such that \cite{arnold}
\begin{equation}
\{Z^A\}\longrightarrow \{\widetilde{Z}^A\}=\{\tilde \Phi, \tilde E ^a, \tilde I ^ a\}\ ,
\end{equation}
\begin{equation}
 \Phi = \tilde \Phi - \delta_{kl} \tilde E ^k \tilde I ^l \ ,\quad
 E^i = - \tilde I ^ {i}, \ \  
E^j = \tilde E ^j,\quad   
 I^{i} = \tilde E ^ i , \ \
 I^j = \tilde I ^j \ ,
 \label{leg}
\end{equation}
where $i\cup j$ is any disjoint decomposition of the set of indices $\{1,...,n\}$,
and $k,l= 1,...,i$. In particular, for $i=\{1,...,n\}$, i.e.,
\begin{equation}
 \Phi = \tilde \Phi - \delta_{ab} \tilde E ^a \tilde I ^b \ ,\quad
 E^a = - \tilde I ^ {a},\quad   
 I^{a} = \tilde E ^ a , 
  \label{tleg}
\end{equation} 
we obtain the total Legendre transformation whereas for $i=\emptyset$, we obtain
the identity transformation, i.e.,
\begin{equation}
 \Phi = \tilde \Phi ,\quad
 E^a =  \tilde E ^ {a},\quad   
 I^{a} = \tilde E ^ a \ . 
  \label{idleg}
\end{equation}  

The main point is now that all the geometric structures to be defined on ${\cal T}$ must be invariant under Legendre transformations in such a way that its $n-$dimensional subspaces ${\cal E}$ inherit this property. First, we introduce the fundamental $1-$form 
\begin{equation}
\Theta = d\Phi - \delta_{ab}I^adE ^b\ , \quad \delta_{ab}={\rm diag}(1,1,\cdots,1)\ ,
\end{equation} 
which is Legendre invariant in the sense that it transforms as $\Theta\rightarrow \tilde \Theta =d\tilde \Phi - \delta_{ab} \tilde I^a d \tilde E^b$.
Notice that the existence of this $1-$form is guaranteed as a consequence of Darboux theorem \cite{her73} because it satisfies the condition
\begin{equation}
\Theta \wedge (d\Theta)^{\wedge n} \neq 0 \ ,
\label{cond}
\end{equation}
which is interpreted as stating that (the tangent space of) ${\cal T}$ is maximally non-integrable. Notice that the $(2n+1)-$form  
$\Theta \wedge (d\Theta)^{\wedge n}=d\Phi \wedge dI^1\cdots \wedge dI^n  \wedge dE^1\cdots \wedge dE^n $ determines the volume 
element $d^{2n+1} Z$  in ${\cal T}$ so that the condition  (\ref{cond}) indicates that the integral is a well-defined operation. 
The pair $({\cal T},\Theta)$ is known as a contact space. 

Let us now introduce a Riemannian metric structure $G=G_{AB}dZ^A dZ^B$ 
in ${\cal T}$. From all possible 
metrics $G$ we choose only those that are invariant under Legendre transformations. Then the triad $({\cal T},\Theta,G)$ determines a
Riemannian contact manifold that is Legendre invariant and is known as the phase manifold. It is now clear that any geometric structures derived from 
$\Theta$ and $G$ under certain conditions can be Legendre invariant. 

Let us now consider the equilibrium manifold ${\cal E}$ as a submanifold of ${\cal T}$ determined by the smooth embedding map 
\begin{equation}
\varphi: {\cal E} \rightarrow {\cal T} \ ,
\end{equation}
or in coordinates 
\begin{equation}
 \varphi :  \{E^a\} \longmapsto \{Z^A(E^a)\}=\{\Phi(E^a), E^a, I^a(E^a)\} \ ,
\label{mapcoo}
\end{equation}
satisfying the condition 
\begin{equation}
\varphi^*(\Theta)=0 \ , \qquad {\rm i.e.,}\qquad \Theta = \delta_{ab} I^a d E^b \ \ {\rm on\ } {\cal E} \ ,
\label{firstinv}
\end{equation}
which is equivalent to the first law as given in Eq.(\ref{first}). As a result of the smoothness of $\varphi$ (and the pullback $\varphi^*)$ and of the invariance of $\Theta$, the first law (\ref{firstinv}) is Legendre invariant in this formalism. 

We can now consider the canonically induced metric
\begin{equation}
g=\varphi^*(G) \Longleftrightarrow g_{ab}= \frac{\partial Z^A}{\partial E^a}\frac{\partial Z^B}{\partial E^b}G_{AB}\ ,  
\end{equation}
as the metric of the equilibrium manifold. In this way, the Riemannian manifold $({\cal E},g)$ can be considered as  invariant in the sense that it is obtained in a canonical manner from the phase manifold $({\cal T},\Theta,G)$ by using only geometric objects that are Legendre invariant. Notice that the definition of the smooth map $\varphi$ as given in Eq.(\ref{mapcoo}) implies that the fundamental equation $\Phi=\Phi(E^a)$ must be known in order to define the equilibrium manifold. This is in accordance with the intuitive description of the equilibrium space given above.

From the above description it follows that the only freedom in the construction of the 
phase manifold is in the choice of the metric $G$. Although Legendre invariance 
implies a series of algebraic conditions for the metric components $G_{AB}$ 
\cite{quev07}, and it can be shown that these conditions are not trivially satisfied, 
the  metric $G$ cannot be fixed uniquely. It is important  to mention that
a straightforward computation shows that the flat 
metric $G=\delta_{AB}dZ^A dZ^B$ is not invariant with respect to the Legendre transformations given 
in Eq.(\ref{leg}).  It then follows that the phase manifold is necessarily curved. 
Moreover, one can show that the Hessian metrics for the equilibrium manifold can be generated from a specific metric 
$G^H$ of the phase manifold according to 
\begin{equation}
g^H=\varphi^*(G^H) = \varphi^*( \Theta^2+\delta_{ab} dE^a dI ^b) = \frac{\partial^2 \Phi} {\partial  E^a \partial E^b} dE^a dE^b \ .
\end{equation}
It is easy to prove that $G^H$ is not invariant under Legendre transformations, indicating that results obtained by using Hessian metrics could depend
on the choice of thermodynamic potential.

If we limit ourselves to the case of total Legendre transformations, 
we find that there exists a class of metrics, 
\begin{equation}
G = \left(d\Phi - I_a dE^a\right)^2  +\Lambda
\left(\xi_{ab}E^{a}I^{b}\right)\left(\chi_{cd}dE^{c}dI^{d}\right) \ , 
\label{gup1}
\end{equation}
parametrized by the diagonal constant tensors $\xi_{ab}$ and $\chi_{ab}$, that are invariant for several choices 
of these free tensors. The constant parameter $\Lambda$ is introduced to guarantee that the units of the first and second
terms coincide. No particular significance is attributed to this constant.

Since the tensors $\xi_{ab}$ and $\chi_{ab}$ must be constant and diagonal, it seems reasonable 
to express them in terms of the usual Euclidean and pseudo-Euclidean metrics 
$\delta_{ab}={\rm diag}(1,...,1)$ and $\eta_{ab} = {\rm diag}(-1, 1, ..., 1)$, respectively. Then, for instance, the choice
$
\xi_{ab}=\delta_{ab}\ , \ \chi_{ab}=\delta_{ab} 
$
corresponds to a Legendre invariant metric which has been used to describe the geometric properties of systems with 
first order phase transitions \cite{quev07,qstv10a}. Moreover, the choice
$
\xi_{ab}=\delta_{ab},\ \chi_{ab}=\eta_{ab}
$
turned out to describe correctly second order phase transitions especially in black hole thermodynamics \cite{qstv10a,aqs08,vqs09,qsv09}. 
At the moment we have no definite explanation for the fact that the signature of $\chi_{ab}$ is able to differentiate between phase transitions of  first and second orders. Nevertheless, it seems that this difference can be used to propose an alternative invariant definition of phase transitions. This task is currently under consideration, and will be presented elsewhere.

Obviously, for a given thermodynamic system it is very important to 
choose the appropriate  metric in order to describe correctly the thermodynamic properties in 
terms of the geometric properties obtained in the context of GTD. In this work, we will consider the van der Waals gas that is characterized by the presence of first order phase transitions. Then, the appropriate metric is
\begin{equation}
G = \left(d\Phi - I_a dE^a\right)^2  +\Lambda
\left(I_aE^{a}\right)\left(dI_bdE^{b}\right) \  ,
\label{gup2}
\end{equation}
from which we can compute the corresponding induced metric
\begin{equation}
g=\varphi^*(G) = \Lambda \left(\frac{\partial \Phi}{\partial E^a} E^a\right)\left(\frac{\partial^2\Phi}{\partial E^b\partial E^c}dE^b dE^c\right)\ .
\label{gdown}
\end{equation}

It is then clear that  once the fundamental equation $\Phi(E^a)$ is given, the expression for the metric of the equilibrium manifold can be found explicitly.

\section{Geometric properties of the equilibrium manifold}
\label{sec:equ}

Let us consider the fundamental equation for the van der Waals gas in the entropy representation
\begin{equation} 
S=\frac{3}{2}\ln\left(U+\frac{a}{V}\right)+\ln(V-b)\ ,
\end{equation}
where $U$ is the internal energy, $V$ is the volume, and $a$ and $b$ are constants. The parameter $b$ is associated with the volume of the molecules of the gas and plays a qualitative role in the description whereas $a$ is responsible for the thermodynamic interaction. The corresponding intensive variables can be obtained from the first law of thermodynamics $dS=\frac{1}{T}dU + \frac{P}{T}dV$ as 
\begin{equation}
T = \left(\frac{\partial S}{\partial U}\right)^{-1} = \frac{2}{3}\left(U+\frac{a}{V}\right)\ ,\quad
P= \left(\frac{\partial S}{\partial V}\right)\left(\frac{\partial S}{\partial U}\right)^{-1}=\frac{2UV^2-aV+3ab}{3V^2(V-b)}\ .
\label{int}
\end{equation}
According to Eq.(\ref{gdown}), in this case the metric 
\begin{equation}
g= \Lambda\left(\frac{\partial S}{\partial U} U +\frac{\partial S}{\partial V}V\right)
 \left( \frac{\partial^2 S}{\partial U^2} dU^2 + 2\frac{\partial^2 S}{\partial U\partial V} dU dV + \frac{\partial^2 S}{\partial V^2} dV^2\right)\ ,
\end{equation}
defines the equilibrium manifold, and for the van der Waals gas reads
\begin{equation}
g=\frac{\Lambda}{2}\frac{5\,U{V}^{2}-3\,UVb-aV+3\,ab}
{(UV+a)^3(V-b)}\left[-\frac{3}{2}V^2dU^2 + 3adUdV -\frac{W(U,V)}{2V^2(V-b)^2} dV^2\right]\ ,
\label{gvdw}
\end{equation}
where
\begin{equation} W(U,V)=2\,{V}^{4}{U}^{2}-2\,{V}^{3}Ua-{a}^{2}{V}^{2}+12\,a{V}^{2}bU+6\,Vb{a}^
{2}-6\,a{b}^{2}UV-3\,{b}^{2}{a}^{2}\ .
\end{equation}
In the limiting case $a=0$ and $b=0$ (ideal gas limit) the above expression reduces to the metric
\begin{equation}
g=-\frac{5\Lambda }{2}\left(\frac{3}{2} \frac{dU^2}{U^2} + \frac{dV^2}{V^2}\right)\ ,
\end{equation}
whose curvature tensor vanishes identically. This proves that the equilibrium manifold of the ideal gas is flat, indicating the lack of thermodynamic interaction.

In the general case $a\neq 0$ and $b\neq 0$, the curvature tensor is different from zero. We interpret this result as an indication of the presence of thermodynamic interaction. The corresponding curvature scalar can be written as
\begin{equation}
R=\frac{N_{vdW}(U,V)}{\left( {V}^{3}U-2\,{V}^{2}a+6\,Vba-3\,{b}^{2}a \right) ^{2} \left( 5
\,U{V}^{2}-3\,UVb-aV+3\,ab \right) ^{3}
}\ ,
\end{equation}
so that the curvature singularities are determined by the zeros of the two polynomials entering the denominator. The function
$N_{vdW}(U,V)$ is a polynomial that is  different from zero at those points where the denominator vanishes. Using the expression for the pressure given in Eq.(\ref{int}), one can show that
\begin{equation}
{V}^{3}U-2\,{V}^{2}a+6\,Vba-3\,{b}^{2}a =\frac{3}{2}(V-b)(PV^3-aV+2ab) 
\label{trans}
\end{equation}
and
\begin{equation}
5\,U{V}^{2}-3\,UVb-aV+3\,ab=3V(V-b)(U+PV)\ . 
\end{equation}
It follows that there exist curvature singularities at those points where the condition $PV^3-aV+2ab=0$ is satisfied. In classical thermodynamics it is known that this condition determines the points where first order phase transitions occur in the van der Waals gas \cite{callen}. As for the second polynomial, one can see that no zeros exist for positive values of the pressure, a condition that is usually assumed as valid in van der Waals systems. 

We conclude that in GTD the equilibrium manifold of the van der Waals gas correctly describes the corresponding thermodynamic behavior.

%%%%%%%%%%%%%%%%%%%%%%%%%%%%%%%%%%%%%%%%%%%%%%%%%%%%%%%%%
%%%%%%%%%%%%%%%%%%%%%%%%%%%%%%%%%%%%%%%%%%%%%%%%%%%%%%%%%%
\section{Geodesic incompleteness}
\label{sec:geo}

In classical thermodynamics, a quasi-static thermodynamic process is a process that happens infinitely slowly and therefore it can be ensured that 
the system passes through a sequence of states that are infinitesimally close to equilibrium and, consequently, the system remains 
in quasi-static equilibrium. 

Since each point of the manifold ${\cal E}$ represents an equilibrium state, a quasi-static process 
can be interpreted as a sequence of points, i. e., as a curve in ${\cal E}$. In particular, the geodesic curves of ${\cal E}$ along which the laws 
of thermodynamics are satisfied can represent 
quasi-static processes that take place inside an isolated system without influence from outside. 

If we introduce the concept of thermodynamic length as $L=\int ds=\int \sqrt{g_{ab}dE^a dE^b}$ in ${\cal E}$, 
the vanishing of the variation $\delta L =0$  leads to the geodesic equation
\begin{equation}
\frac{d ^2E^a}{d\tau^2} + 
\Gamma^a_{\ bc} \frac{dE^b}{d\tau} \frac{dE^c}{d\tau} = 0 \ ,
\label{geo1}
\end{equation}
where $\Gamma^a_{\ bc}$ are the Christoffel symbols of the thermodynamic metric $g$, and $\tau$ is an arbitrary affine parameter
along the geodesic. 

\begin{figure}
\includegraphics[height=.3\textheight]{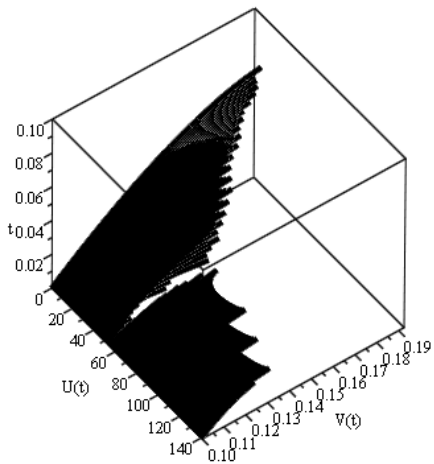}
\caption{Geodesics in the equilibrium manifold of the van der Waals gas with initial value $V(\tau=0)=0.1$, $\dot U (\tau =0)=0$, $\dot V (\tau =0)=1$, and different initial values $U(\tau=0)$ in the range $[0,140]$.  }
\label{fig1}
\end{figure}

In the case of the equilibrium manifold of the van der Waals gas, the geodesic equations are
\begin{equation}
 \frac{d^2 U}{d \tau^2} + \Gamma^U_{UU} \left(\frac{dU}{d\tau}\right)^2+2\Gamma^U_{UV}\frac{dU}{d\tau}\frac{dV}{d\tau}+
\Gamma^U_{VV} \left(\frac{dV}{d\tau}\right)^2=0\ ,
\end{equation}
\begin{equation}
\frac{d^2 V}{d \tau^2} + \Gamma^V_{UU} \left(\frac{dU}{d\tau}\right)^2+2\Gamma^V_{UV}\frac{dU}{d\tau}\frac{dV}{d\tau}+
\Gamma^V_{VV} \left(\frac{dV}{d\tau}\right)^2=0\ ,
\end{equation}
where the explicit form of the Christoffel symbols can be calculated from Eq.(\ref{gvdw}). The resulting differential equations
are highly non-trivial and cannot be treated analytically. Instead, we perform a numerical analysis for a large number of different 
initial conditions. The behavior of the geodesics is illustrated in Fig.\ref{fig1}. The range of the initial conditions was chosen 
such that the geodesics always reach a maximum value for $V(\tau)$. 
%%%%%%%%%%
%%%%%%%%%

\begin{figure}
\includegraphics[height=.3\textheight]{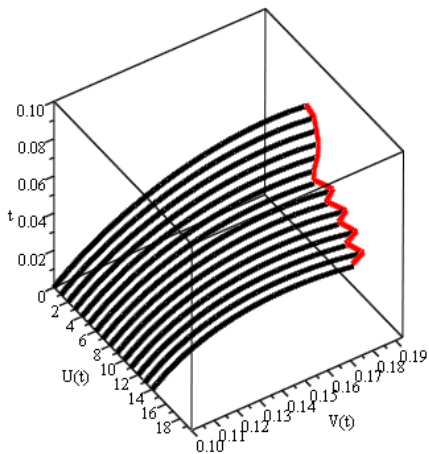}
\caption{Geodesics in the equilibrium manifold of the van der Waals gas for different initial values $U(\tau=0)$ and $V(\tau=0)=0.1$. The incompleteness of the geodesics occur at the points where first order phase transitions occur.  }
\label{fig2}
\end{figure}

To clarify this point we perform an analysis in a smaller range $V(\tau=0)\in [0,14]$ as depicted in Fig.\ref{fig2}.
%%%%%%%%%%%%%%
The main observation is that the geodesics are incomplete, i.e, there exists a maximum value 
of the affine parameter $\tau_{max}$ for which the numerical integration delivers an end 
value of $U_{max}$ and $V_{max}$. After this point the geodesics are not defined.
To understand the origin of this behavior, we analyze numerically the end points 
 $U_{max}$ and $V_{max}$ and found that they satisfy the relationship
\begin{equation}
V_{max}^{3}U_{max}-2\,V_{max}^{2}a+6\,V_{max}ba-3\,{b}^{2}a = 0 \ ,
\end{equation} 
which, according to Eq.(\ref{trans}), determines the points where the curvature of the equilibrium manifold diverges and first order phase transitions take place.
It then follows that  the geodesic incompleteness is due
to the appearance of phase transitions. This result corroborates the fact that curvature singularities can also be defined by using the concept of geodesic incompleteness (see, for instance, \cite{hawellis}). We conclude that in GTD we can also use the geodesics to find the points where phase transitions can occur.

%%%%%%%%%%%%%%%%%%%%%%%%%%%%%%%%%%%%%%%%%%
%%%%%%%%%%%%%%%%%%%%%%%%%%%%%%%%%%%%%%%%%%
\section{Conclusions}
\label{sec:con}

In this work, we presented the formalism of GTD starting from the geometric properties of the equilibrium manifold and by demanding that they do not depend on the choice of thermodynamic potential. In this manner, it becomes clear why it is necessary to introduce the auxiliary phase manifold where Legendre transformations are well-defined. Since the equilibrium manifold is defined by means of a smooth map  as a subspace of the phase manifold that is manifestly Legendre invariant, it inherits the invariance property. 

We use a particular metric of the phase manifold, which is is invariant under total Legendre transformations, to compute the induced metric for the equilibrium manifold. It is shown that all the geometric structure of the equilibrium manifold is determined by the fundamental equation only. 

The case of a thermodynamic system described by the van der Waals fundamental equation is analyzed in detail. First, we found the explicit form of the metric of the equilibrium manifold, and showed that it corresponds in general to a curved space, indicating that the curvature can be used as a measure of the thermodynamic interaction. In the limiting case of an ideal gas, the curvature vanishes as expected for a system with no thermodynamic interaction. The curvature singularities of the van der Waals equilibrium manifold were shown to be located only at those places where first order phase transitions occur. This, of course, is an indication of the break down of the equilibrium thermodynamics approach and, consequently, of GTD.    

We investigated numerically the geodesic equations of the van der Waals equilibrium manifold. It was shown that for a certain range of initial values, the geodesics cannot be extended after a particular value of the affine parameter. The points where this geodesic incompleteness was detected turned out to coincide with the points where the curvature diverges. This is in accordance with the result known in differential geometry about the equivalence between curvature singularities and geodesic incompleteness. We interpret this result in GTD as an additional criterion for detecting phase transitions in the equilibrium manifold.

%%%%%%%%%%%%%%%%%%%%%%%%%%%%%%%%%%%%%%%%%%%%%%%%
%% BACKMATTER
%%%%%%%%%%%%%%%%%%%%%%%%%%%%%%%%%%%%%%%%%%%%%%%%

\begin{acknowledgments}
It is a great pleasure to dedicate this work to Professor Mario Novello on the occasion of his 70-th birthday. 
We would like to thank the members of the GTD-group at the UNAM for stimulating discussions and interesting comments.
One of us (HQ) would like to thank the faculty members and students of the Department of Physics for excellent hospitality during his stay 
at the Al-Farabi Kazakh National University where part of this work was done.
This work was partially supported by DGAPA-UNAM, grant No IN106110, and Conacyt, grant No. 166391. 
\end{acknowledgments}



\bibliographystyle{aipproc}   % if natbib is available

%\bibliography{sample}

\IfFileExists{\jobname.bbl}{}
 {\typeout{}
  \typeout{******************************************}
  \typeout{** Please run "bibtex \jobname" to optain}
  \typeout{** the bibliography and then re-run LaTeX}
  \typeout{** twice to fix the references!}
  \typeout{******************************************}
  \typeout{}
 }

%\end{document}

%%%%%%%%%%%%%%%%%%%%%%%%%%%%%%%%%%%%%%%%%%%
%% The following lines show an example how to produce a bibliography
%% without the help of the BibTeX program. This could be used instead
%% of the above.
%%%%%%%%%%%%%%%%%%%%%%%%%%%%%%%%%%%%%%%%%%%

\end{document}